\documentclass[english,aps,preprint,showpacs]{revtex4}
\usepackage[T1]{fontenc}
\usepackage[latin9]{inputenc}
\usepackage{graphicx}

\providecommand{\tabularnewline}{\\}

\usepackage{babel}

\begin{document}

\title{Search for Top Quark FCNC Couplings in $Z'$ Models at the LHC and CLIC}

\author{O. \c{C}ak\i{}r}

\email{ocakir@science.ankara.edu.tr}

\affiliation{Ankara University, Faculty of Sciences, Department of Physics, 06100,
Tandogan, Ankara, Turkey}

\author{I.T. \c{C}ak\i{}r}

\email{tcakir@mail.cern.ch}

\affiliation{Physics Department, CERN, 1211, Geneva 23, Switzerland}

\author{A. Senol}

\email{asenol@kastamonu.edu.tr}

\affiliation{Kastamonu University, Faculty of Arts and Sciences, Department of
Physics, 37100, Kuzeykent, Kastamonu, Turkey}

\author{A.T. Tasci}

\email{atasci@kastamonu.edu.tr}

\affiliation{Kastamonu University, Faculty of Arts and Sciences, Department of
Physics, 37100, Kuzeykent, Kastamonu, Turkey}

\begin{abstract}
The top quark is the heaviest particle to date discovered, with a
mass close to the electroweak symmetry breaking scale. It is expected
that the top quark would be sensitive to the new physics at the TeV
scale. One of the most important aspects of the top quark physics
can be the investigation of the possible anomalous couplings. Here,
we study the top quark flavor changing neutral current (FCNC) couplings
via the extra gauge boson $Z'$ at the Large Hadron Collider (LHC)
and the Compact Linear Collider (CLIC) energies. We calculate the
total cross sections for the signal and the corresponding Standard
Model (SM) background processes. For an FCNC mixing parameter $x=0.2$
and the sequential $Z'$ mass of $1$ TeV, we find the single top
quark FCNC production cross sections $0.38(1.76)$ fb at the LHC with
$\sqrt{s_{pp}}=7(14)$ TeV, respectively. For the resonance production
of sequential $Z'$ boson and decays to single top quark at the Compact
Linear Collider (CLIC) energies, including the initial state radiation
and beamstrahlung effects, we find the cross section $27.96(0.91)$
fb at $\sqrt{s_{e^{+}e^{-}}}=1(3)$ TeV, respectively. We make the
analysis to investigate the parameter space (mixing-mass) through
various $Z'$ models. It is shown that the results benefit from the
flavor tagging.
\end{abstract}


\maketitle

\section{introduction}

The top quark is a wonderful probe for the new physics beyond the
Standard Model (SM) via its decays and productions at high energy
colliders. At the Large Hadron Collider (LHC) the top quark events
will be produced copiously. We may anticipate the discovery of new
physics by observing the anomalous couplings in the top quark
sector. The flavour changing neutral currents (FCNC) couplings of
the top quark can also be enhanced to observable levels in some new
physics models. Many extensions of the SM predict the extra gauge
bosons, especially the $Z'$-boson has been the object of extensive
phenomenologial studies (\cite{langacker08} and references therein).
An extra $U(1)$ gauge boson $Z'$ can induce flavor changing neutral
currents. In the models with an extra $U(1)$ group the $Z'$ boson
can have tree-level or an effective $Z'-q-q'$ couplings, where $q$
and $q'$ are both the up-type quarks or down-type quarks. The LHC
can probe some parameter space of the $Z'$ models provided necessary
luminosity. If a $Z'$ boson is found at the LHC, the underlying
model could be best identified at the linear colliders through the
polarization observables.

The current experimental searches of the $Z'$ boson from Drell-Yan
cross sections at Tevatron have put lower limits on the mass range
$0.6-1.0$ TeV at $95\%$ C.L. depending on the specific $Z'$ models
\cite{Amsler08}. From the electroweak precision data analysis, the
improved lower limits on the $Z'$ mass are given in the range $1.1-1.4$
TeV at $95\%$ C.L. \cite{Erler09}. These limits on the $Z'$ boson
mass favors higher energy ($\geq1$ TeV) collisions for direct observation
of the signal. It is also possible that the $Z'$ bosons can be much heavy
or weak enough to escape beyond the discovery reach expected at the
LHC. In this case, only the indirect signatures of $Z'$ exchanges
may occur at the high energy colliders.

The $Z'$ models have some special names \cite{Amsler08}: sequential
$Z'_{S}$ model has the same couplings to the fermions as that of
the $Z$ boson of the SM, left-right symmetric $Z'_{LR}$ model has
the couplings a combination of right-handed and $B-L$ neutral currents,
the $Z'_{\psi}$, $Z'_{\chi}$ and $Z'_{\eta}$ models corresponding
to the specific values of the mixing angle in the $E_{6}$ model have
different couplings to the fermions.

A work which addresses the effects of tree-level FCNC interactions 
induced by an additional $Z'$ boson on the single top quark production at 
the LHC ($\sqrt{s}=14$ TeV) and International Linear Collider 
(ILC) ($\sqrt{s}=1$ TeV) has been performed in Ref. \cite{arhrib06}. 
In the paper, the relevant signal cross sections have been calculated and 
especially the charm tagging to identify the signal has been discussed. 
Even though it shows the potential of the LHC and ILC for a 
certain parameter set in the single top FCNC production, 
it is free of a detailed analysis 
for the background, including Monte Carlo (MC) simulation 
and observability for a full parameter space. 

In our work, we investigate both the single and pair production
of top quarks via FCNC interactions through $Z'$ boson exchange at
the LHC and Compact Linear Collider (CLIC) \cite{clic}. 
The aim of this paper is to complement the previous other works by 
studying the signal and background in detail in the 
same MC framework. Therefore,
we implement the related interaction 
vertices into the MC software, and study the FCNC parameters in detail
as well as the effects of initial state radiation (ISR) and 
beamstrahlung (BS) in the $e^+e^-$ collisions. 
Another feature of our work is that we analyze the
signal observability (via contour plots) 
for the $Z'$ boson and top quark FCNC interactions.

In section II, we calculate the decay widths of $Z^{\prime}$ boson for the mass
range 1000-3000 GeV in the framework of the model 
which has already been detailed in Ref. \cite{arhrib06}.
An analysis of the mass and coupling
strength parameter space for different $Z^{\prime}$ models are given for the
single and pair production of top quarks at the LHC in section III
and at the CLIC in section IV. Taking into account the initial state
radiation (ISR) and beamstrahlung (BS) effects in the $e^{+}e^{-}$
collisions, we analyzed the signal observability for the $Z'$ boson
and top quark FCNC interactions. In order to enrich
the signal statistics even at the small couplings we consider both
$t\bar{c}$ and $\bar{t}c$ single top productions in the final state.
The analysis for the signal significance and 
conclusion are given in sections V and VI, respectively.

\section{Model}

In the gauge eigenstate basis, following the formalism given in Ref.
\cite{arhrib06,langacker00,cheung07}, the additional neutral current Lagrangian associated
with the $U(1)^{'}$ gauge symmetry can be written as\begin{equation}
\mathcal{L}'=-g'\sum_{f,f'}\bar{f}\gamma^{\mu}\left[\epsilon_{L}^{'}(ff')P_{L}+\epsilon_{R}^{'}(ff')P_{R}\right]f'Z_{\mu}^{'}\end{equation}
 where $\epsilon_{L,R}^{'}(ff')$ are the chiral couplings of $Z'$ boson
with fermions $f$ and $f'$. The $g'$ is the gauge coupling of the
$U(1)^{'}$, and $P_{R,L}=(1\pm\gamma^{5})/2.$ Here, we assume that
there is no mixing between the $Z$ and $Z'$ bosons as favored by
the precision data. Flavor changing neutral currents (FCNCs) arise
if the chiral couplings are nondiagonal matrices. In case the $Z'$
couplings are diagonal but nonuniversal, flavor changing couplings
are emerged by fermion mixing. In the interaction basis the FCNC for
the up-type quarks are given by

\begin{equation}
\mathcal{J}_{FCNC}^{u'}=\left(\overline{u},\overline{c},\overline{t}\right)\gamma_{\mu}(\epsilon_{L}^{u'}P_{L}+\epsilon_{R}^{u'}P_{R})\left(\begin{array}{c}
u\\
c\\
t\end{array}\right)\end{equation}
 where the chiral couplings are given by \cite{arhrib06} 

\begin{equation}
\epsilon_{L}^{u'}=C_{L}^{u}\left(\begin{array}{ccc}
1 & 0 & 0\\
0 & 1 & 0\\
0 & 0 & x\end{array}\right)\quad\mbox{and}\quad\epsilon_{R}^{u'}=C_{R}^{u}\left(\begin{array}{ccc}
1 & 0 & 0\\
0 & 1 & 0\\
0 & 0 & 1\end{array}\right) .
\end{equation}

In general, the effects of these FCNCs may occur both in the up-type
sector and down-type sector after diagonalizing their mass matrices.
For the right-handed up-sector and down-sector one assumes that the
neutral current couplings to $Z'$ are family universal and flavor
diagonal in the interaction basis. In this case, unitary rotations
($V_{L,R}^{f}$) can keep the right handed couplings flavor diagonal,
and left handed sector becomes nondiagonal. The chiral couplings of
$Z'$ in the fermion mass eigenstate basis are given by 

\begin{equation}
B_{L}^{ff'}\equiv V_{L}^{f}\epsilon_{L}^{'}(ff')V_{L}^{f^{\dagger}}\quad\mbox{and}\quad B_{R}^{ff'}\equiv V_{R}^{f}\epsilon_{R}^{'}(ff')V_{R}^{f^{\dagger}}\end{equation}
 here the CKM matrix can be written as $V_{CKM}=V_{L}^{u}V_{L}^{d\dagger}$
with the assumption that the down-sector has no mixing. The flavor
mixing in the left-handed quark fields is simply related to $V_{CKM}$, 
assuming the up sector diagonalization 
and unitarity of the CKM matrix one can find the couplings \cite{arhrib06}

\begin{equation}
B_{L}^{u}\equiv V_{CKM}^{\dagger}\epsilon_{L}^{u'}V_{CKM}\approx\left(\begin{array}{ccc}
1 & (x-1)V_{ub}V_{cb}^{*} & (x-1)V_{ub}V_{tb}^{*}\\
(x-1)V_{cb}V_{ub}^{*} & 1 & (x-1)V_{cb}V_{tb}^{*}\\
(x-1)V_{tb}V_{ub}^{*} & (x-1)V_{tb}V_{cb}^{*} & x\end{array}\right) .
\end{equation}

The FCNC effects from the $Z'$ mediation have been studied for the
down-type sector and implications in flavor physics
\cite{Leroux:2001fx,barger04,Barger:2004qc,Barger:2004hn,Chen:2006vs,He:2006bk,Chiang:2006we,Baek:2006bv,cheung07,barger09,barger09-1}
and up-type sector in top quark production
\cite{arhrib06,CorderoCid:2005kp,Yue:2003wd,Lee:2000km,Yue:2006qd,delAguila:2009gz}.
The parameters for different $Z'$ models are given in Table
\ref{tab:1}. In numerical calculations, we take the coupling
$g'\simeq0.65$ for the sequential model and $g'\simeq0.40$ for other
models. In the left-right symmetric model we use $g_{L}=g_{R}$, and
the chiral couplings $C_{L}^{f}=-\sqrt{3/5}(B-L)/2\alpha_{LR}$ and
$C_{R}^{f}=\sqrt{3/5}(\alpha_{LR}I_{3R}^{f})+C_{L}^{f}$ where the
parameter $\alpha_{LR}\simeq1.52$. Here, the $B$ and $L$ denote the
baryon and lepton numbers of the corresponding fermion.

\begin{table}
\caption{The chiral couplings for different $Z'$ models.\label{tab:1}}

\begin{tabular}{|c|c|c|c|c|c|}
\hline
 & $Z'_{S}$  & $Z'_{LR}$  & $Z'_{\chi}$  & $Z'_{\psi}$  & $Z'_{\eta}$\tabularnewline
\hline
$C_{L}^{u}$  & 0.3456  & -0.08493  & $-1/2\sqrt{10}$  & $1/\sqrt{24}$  & $-1/\sqrt{15}$\tabularnewline
\hline
$C_{R}^{u}$  & -0.1544  & 0.5038  & $1/2\sqrt{10}$  & $-1/\sqrt{24}$  & $1/\sqrt{15}$\tabularnewline
\hline
$C_{L}^{d}$  & -0.4228  & -0.08493  & $-1/2\sqrt{10}$  & $1/\sqrt{24}$  & $-1/\sqrt{15}$\tabularnewline
\hline
$C_{R}^{d}$  & 0.0772  & -0.6736  & $-3/2\sqrt{10}$  & $-1/\sqrt{24}$  & $-1/2\sqrt{15}$\tabularnewline
\hline
$C_{L}^{e}$  & -0.2684  & 0.2548  & $3/2\sqrt{10}$  & $1/\sqrt{24}$  & $1/2\sqrt{15}$\tabularnewline
\hline
$C_{R}^{e}$  & 0.2316  & -0.3339  & $1/2\sqrt{10}$  & $-1/\sqrt{24}$  & $1/\sqrt{15}$\tabularnewline
\hline
$C_{L}^{\nu}$  & 0.5  & 0.2548  & $3/2\sqrt{10}$  & $1/\sqrt{24}$  & $1/2\sqrt{15}$\tabularnewline
\hline
\end{tabular}
\end{table}

\begin{figure}
\includegraphics[scale=0.8]{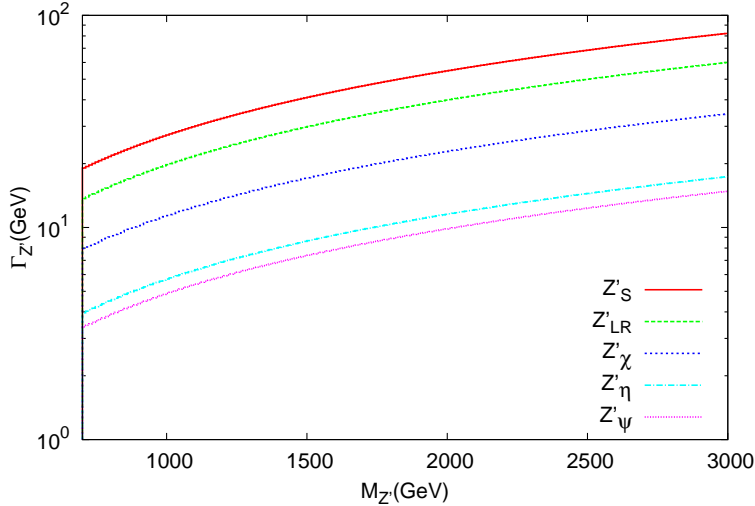}

\caption{The decay widths of $Z'$ boson depending on its mass for different
models with the FCNC parameter $x=0.2$. \label{fig:1}}

\end{figure}

For numerical calculations we have implemented the $Z^{\prime}-q-q'$
interaction vertices into the CompHEP package \cite{Pukhov:1999gg}.
The decay widths of $Z'$ boson for different mass values in the sequential
model, $LR$ symmetric model and $E_{6}$-inspired models are given
in Table \ref{tab:2}. For the parameter $x=1$, both the left-handed
and right-handed couplings become universal, and family diagonal.
In this case we cannot see the FCNC effects on the decay widths and
cross sections. In order to include a little from these effects on
the decay width, we take the parameter $x=0.2$ as shown in Fig. \ref{fig:1}.
The effect of this FCNC reduces the decay width at most $10\%$ for
the sequential model in the relevant mass range. 
The decay widths are compared with the similar results 
from Ref. \cite{arhrib06} for $x=0.1$ to prove the implementation.

\begin{table}
\caption{The $Z'$ boson decay widths (in GeV) for different mass values in
the various models with $x=0.2$ (where the numbers in the paranthesis
denotes the values for $x=1$).\label{tab:2}}

\begin{tabular}{|c|c|c|c|c|c|}
\hline
$M_{Z'}$(GeV)  & $Z'_{S}$  & $Z'_{LR}$  & $Z'_{\chi}$  & $Z'_{\psi}$  & $Z'_{\eta}$\tabularnewline
\hline
1000  & 27.36(29.49)  & 20.08(20.09)  & 11.39(11.55)  & 4.90(5.16)  & 5.73(6.14)\tabularnewline
\hline
1500  & 41.13(44.65)  & 30.30(30.35)  & 17.11(17.38)  & 7.39(7.84)  & 8.65(9.38)\tabularnewline
\hline
2000  & 54.87(59.72)  & 40.49(40.58)  & 22.83(23.21)  & 9.87(10.50)  & 11.57(12.58)\tabularnewline
\hline
2500  & 68.62(74.76)  & 50.66(50.78)  & 28.54(29.03)  & 12.35(13.16)  & 14.47(15.77)\tabularnewline
\hline
3000  & 82.35(89.79)  & 60.82(60.98)  & 34.26(34.85)  & 14.82(15.81)  & 17.38(18.96)\tabularnewline
\hline
\end{tabular}
\end{table}

\section{Proton-proton collisions}

\subsection{The single top quark production}

\begin{figure}
\includegraphics[scale=0.48]{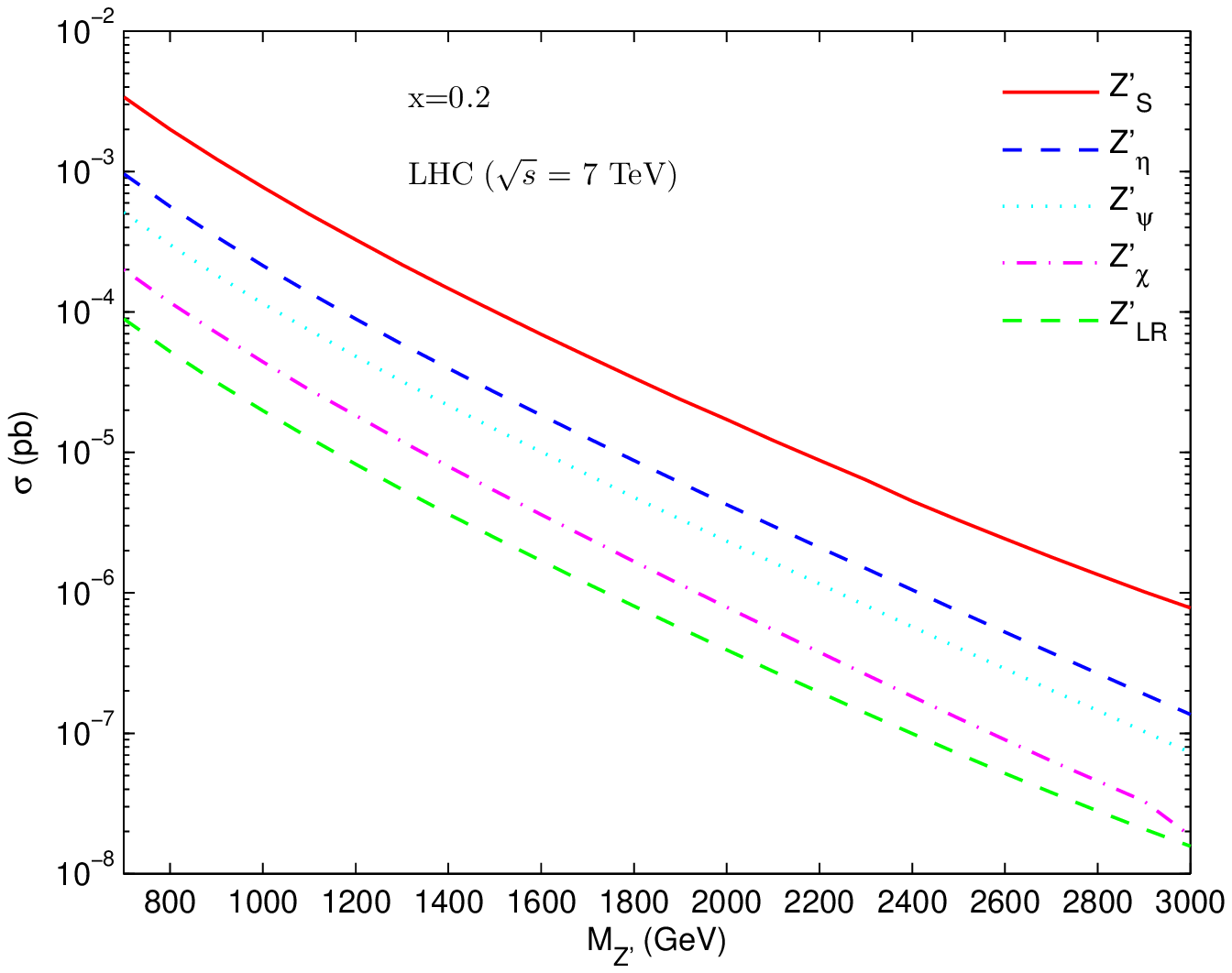} \includegraphics[scale=0.48]{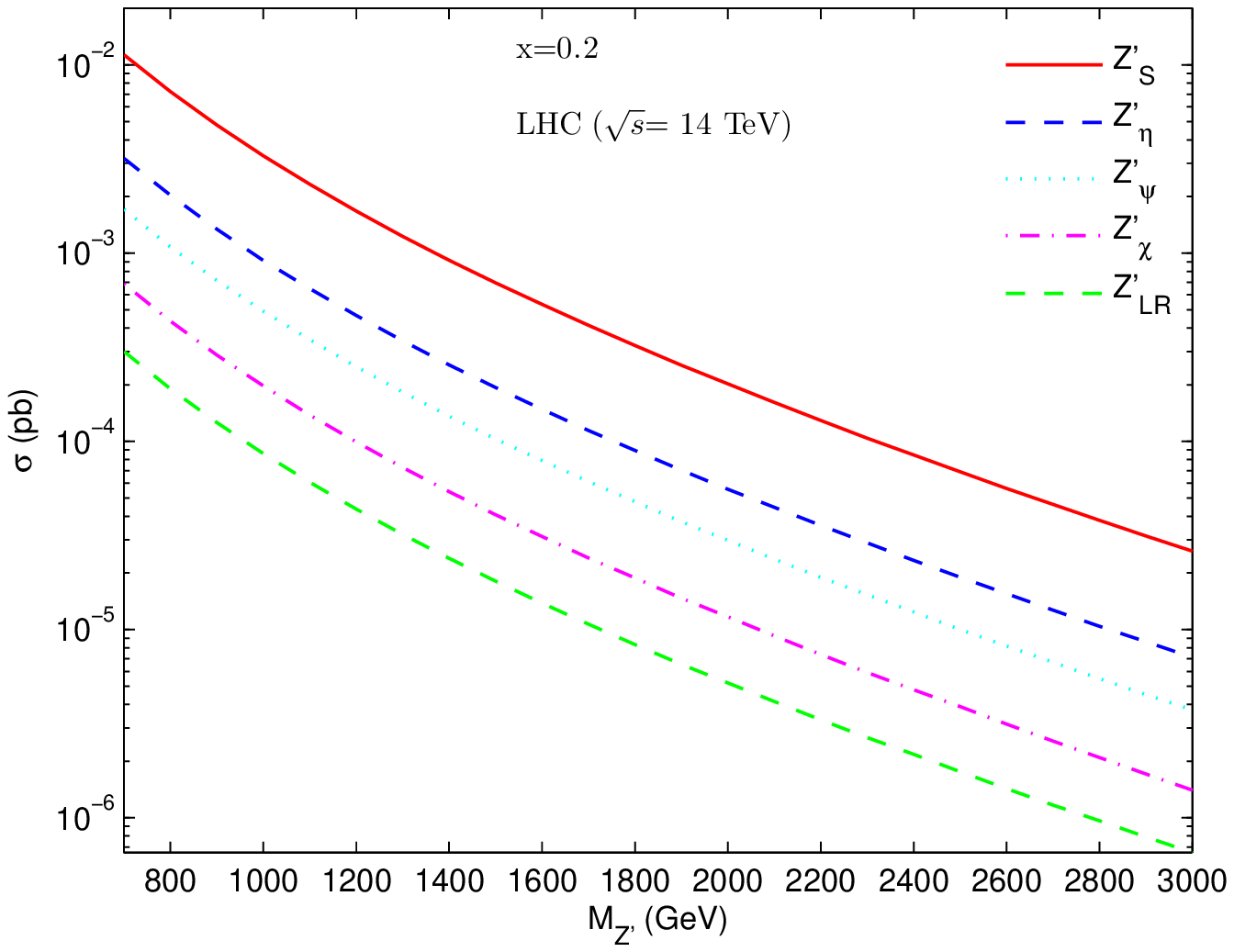}

\caption{The cross sections for $pp\to(t\bar{c}+\bar{t}c)X$ versus the $Z'$
boson mass at the LHC ($\sqrt{s}=$7 TeV left and 14 TeV right). The
lines are for the five $Z'$ models explained in the text. \label{fig:2}}

\end{figure}

The cross sections for the process $pp\to(t\bar{c}+\bar{t}c)X$ depending
on the $Z'$ boson mass at the LHC (both for 7 TeV and 14 TeV) are
given in Fig. \ref{fig:2} by using parton distribution
function library CTEQ6L \cite{Pumplin:2002vw}. Here, the $Z'$
boson contributes through the $s$- and $t$-channel diagrams, and
the associated production of single top quarks in the final state
$t\bar{c}$ well dominates over the $tc$ final state. For this process
the cross section at $\sqrt{s}=14$ TeV is about 3 times larger than
the case at $\sqrt{s}=7$ TeV. Here, we also check our results for 
$x=0.1$ which agrees with that of the results of Ref. \cite{arhrib06}.

\begin{figure}
\includegraphics[scale=0.7]{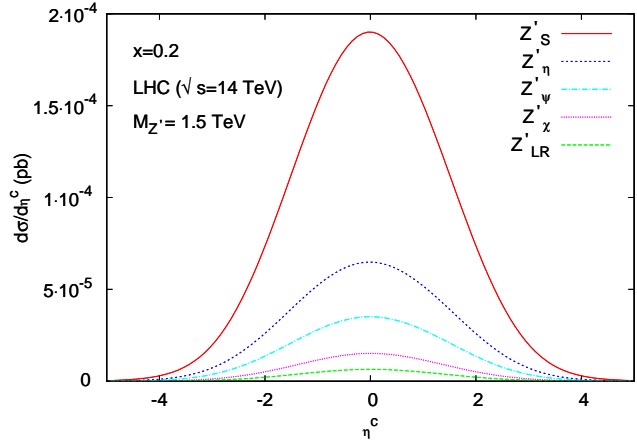}

\caption{The rapidity distributions of the charm quarks ($c$ and $\bar{c}$)
at the LHC with the center of mass energy of 14 TeV. Here, we take
the mass $M_{Z'}=1.5$ TeV and the FCNC parameter $x=0.2$.\label{fig:3}}

\end{figure}

\begin{figure}
\includegraphics[scale=0.7]{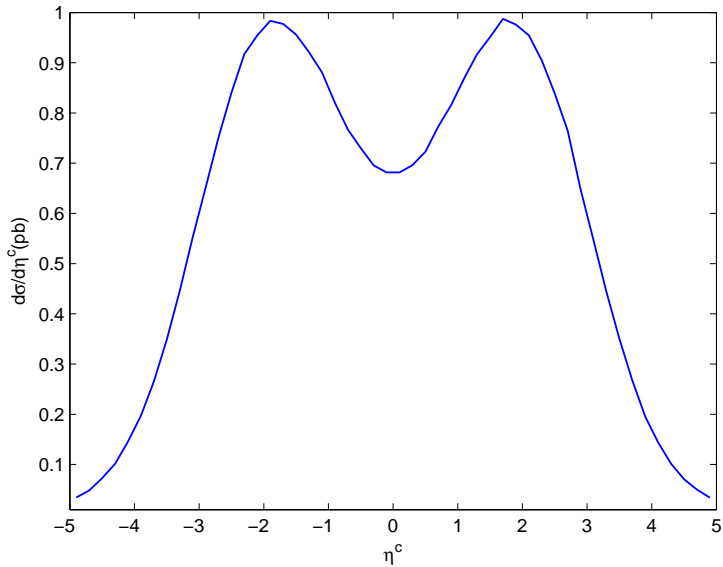}

\caption{The rapidity distribution of the charm quarks ($c$ and $\bar{c}$)
for the background process ($pp\to(t\bar{c}+\bar{t}c)X$) at the LHC
with $\sqrt{s}=14$ TeV. \label{fig:4}}

\end{figure}

The rapidity distribution of the charm quarks ($c$ and $\bar{c}$)
from the signal are shown in Fig. \ref{fig:3} at the collision energy
of 14 TeV. We sum up the $c$ and $\bar{c}$ distributions, since
there will be no clear difference for them. There is a peak
in the $c$-quark rapidity distribution $\eta^{c}\simeq 0$ with
the tails extending to $|\eta^{c}|\simeq2.5$. We may apply a rapidity
cut $|\eta^{j}|<2.5$ for the signal and background analysis. \textit{\emph{The
rapidity distribution of final state $c$ quarks in the background
process $pp\to(t\bar{c}+\bar{t}c)X$ is shown in Fig.}} \ref{fig:4}.

\begin{figure}
\includegraphics[scale=0.7]{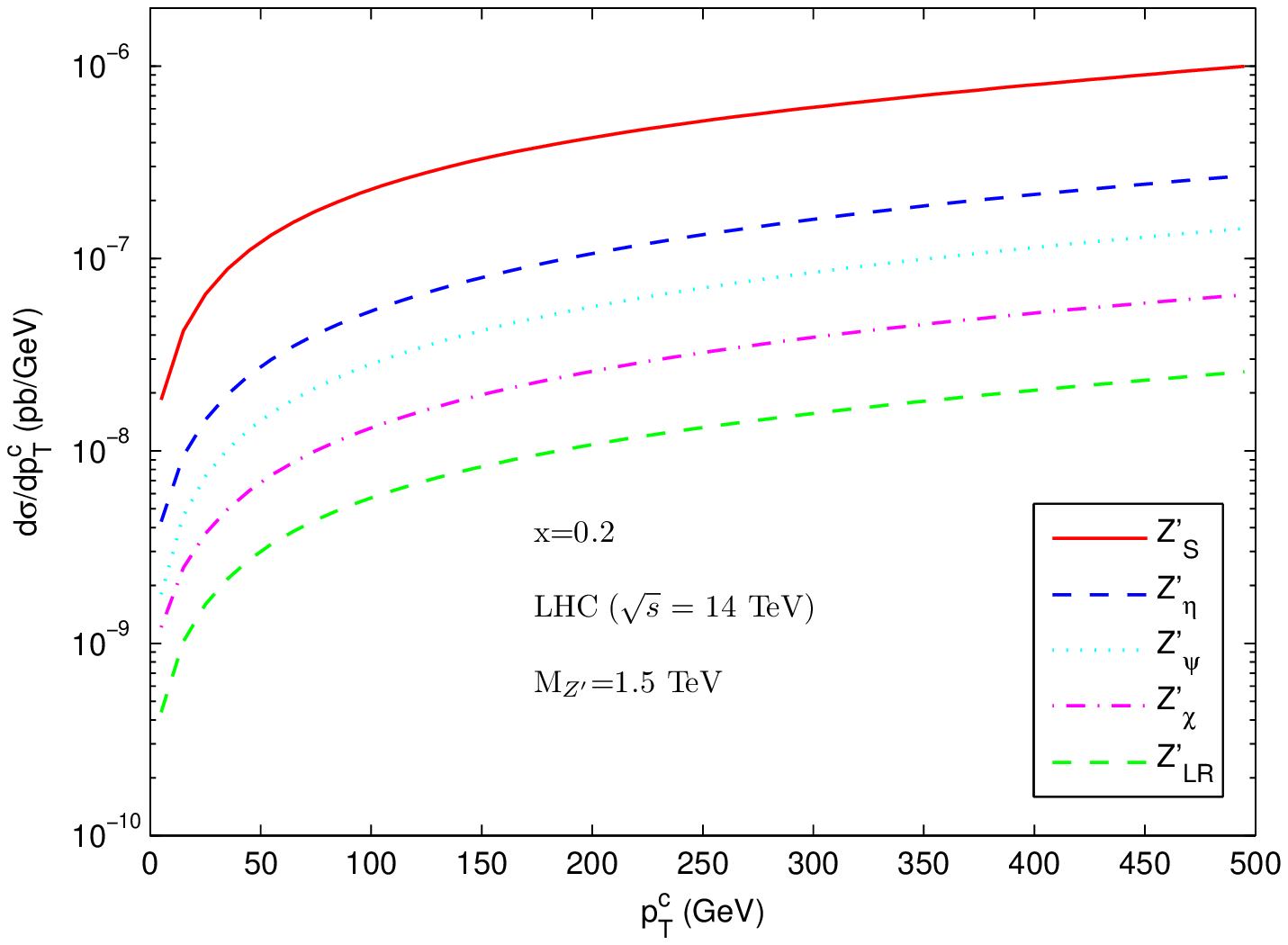}

\caption{The $p_{T}$ distribution of the charm quark for the signal process
$pp\to t\bar{c}+\bar{t}$cX at the LHC ($\sqrt{s}=14$ TeV) with the
FCNC parameter x=0.2. \label{fig:5}}

\end{figure}

\begin{figure}
\includegraphics[scale=0.7]{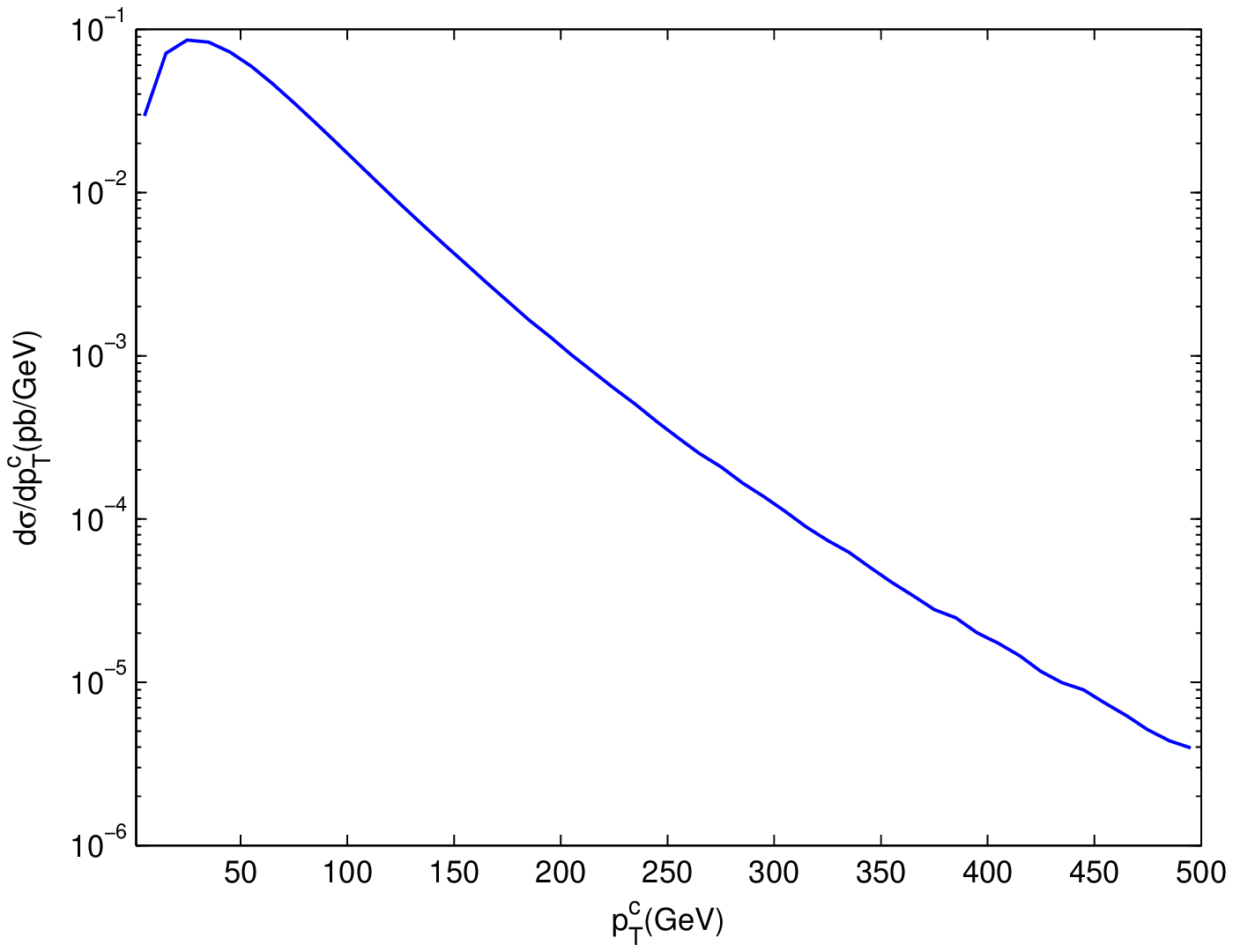}

\caption{The $p_{T}$ distribution of the charm quarks ($c$ and $\bar{c}$)
for the background process ($pp\to t\bar{c}+\bar{t}$cX) at the LHC
with $\sqrt{s}=14$ TeV). \label{fig:6} }

\end{figure}

Fig. \ref{fig:5} shows the $p_{T}$ distributions of the $c$-quark
in the signal process with $M_{Z^{'}}$=1.5 TeV for the parameter
$x=0.2$ at the $pp$ center of mass energy of 14 TeV. A high $p_{T}$
cut ($p_{T}>M_{Z'}/2-4\Gamma_{Z'}$) reduces the background significantly
without affecting much the signal cross section in the interested
$Z'$ mass range. The $p_{T}$ distribution of the $c$-quark from
the background is shown in Fig. \ref{fig:6}. We may apply these cuts
to make analyses with the signal and background.

\begin{figure}
\includegraphics[scale=0.8]{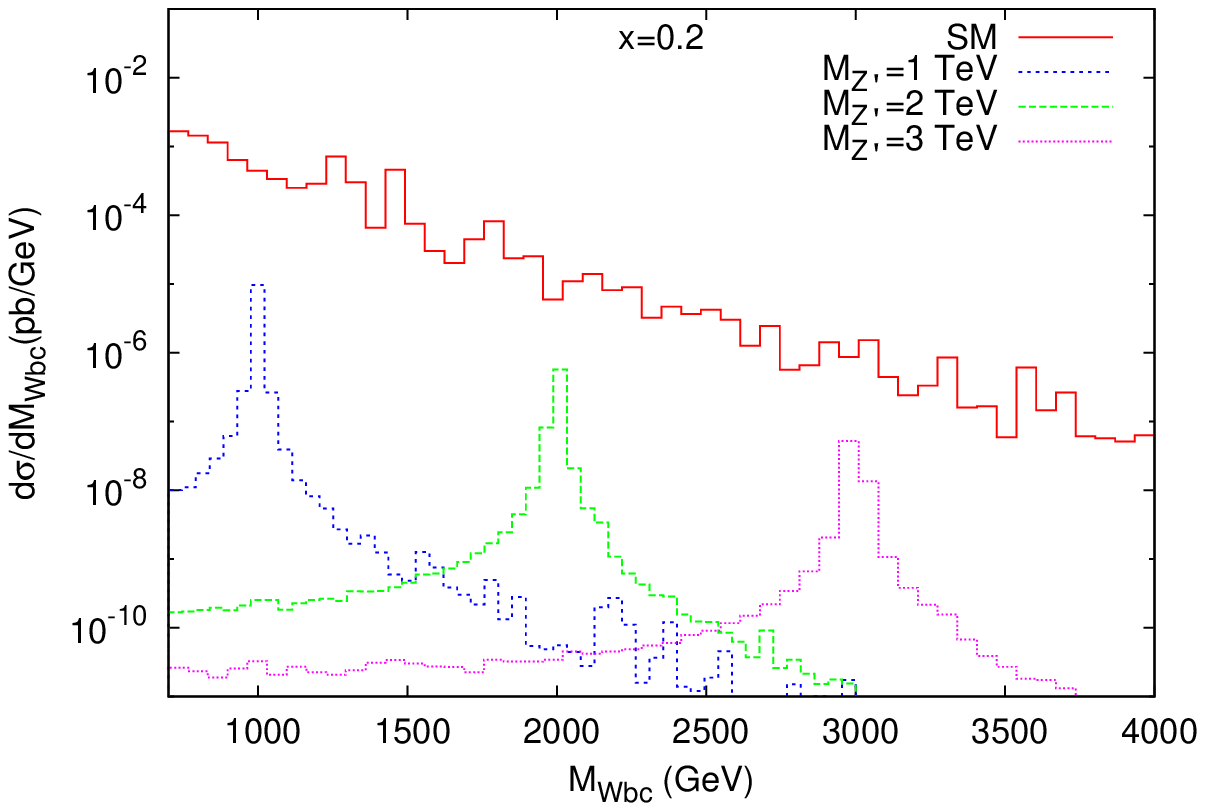}

\caption{The invariant mass distribution of the $W^{+}b\bar{c}$ system for
the SM and $Z'$ sequential model with different mass values of $Z'$
at the LHC with $\sqrt{s}=14$ TeV. \label{fig:7} }

\end{figure}

We plot the invariant mass distribution of the $W^{+}b\bar{c}$ system
for the signal (sequential model with $x=0.2$ and $M_{Z'}=1$, $2$
and $3$ TeV) and background at the LHC with $\sqrt{s}=14$ TeV in
Fig. \ref{fig:7}.

\textit{\emph{Here, we consider two types of backgrounds for the analysis.
The first one has the same final state ($t\bar{c}$ and $\bar{t}c$)
as expected for the signal processes and the other one (single top
associated with a $b$-jet) is the irreducible background and contributes
to the similar final state, assuming the $b$- quark which may be
misidentified as the charmed jet.}} One can apply very high transverse
momentum ($p_{T}$) cut for the $b$-jets and $c$-jets. Employing
the variable $p_{T}$ cuts such that $p_{T}>M_{Z'}/2-4\Gamma_{Z'}$
for different $Z'$ mass values and the rapidity cuts $|\eta|<2.5$
for the central detector coverage, \textit{\emph{in Table \ref{tab:tab3},
we give the statistical significance (SS) values by using, \begin{eqnarray}
SS & = & \sqrt{2L_{int}\epsilon[(\sigma_{S}+\sigma_{B})\ln(1+\sigma_{S}/\sigma_{B})-\sigma_{S}]}\end{eqnarray}
 where $\sigma_{S}$ and $\sigma_{B}$ denotes signal and background
cross sections in the invariant mass interval of $M_{Z'}-2\Gamma_{Z'}<M_{tc}<M_{Z'}+2\Gamma_{Z'}$
by assuming integrated luminosity of $L_{int}=10^{5}$ pb$^{-1}$
per year. For the $Z'$ coupling parameter $x=0.1$, the LHC is able
to measure the $Z'$ mass up to about 2 TeV with single top FCNC.}}

\begin{table}
\caption{The signal significance ($SS$) for the $Z'$ boson predicted in different
models, the $SS$ values correspond to the case $x=0.1$. The integrated
luminosity is taken to be $L_{int}=10^{5}$pb$^{-1}.$ \label{tab:tab3}}

\begin{tabular}{|c|c|c|c|c|c|}
\hline
$M_{Z'}$ (GeV)  & $Z'_{S}$  & $Z'_{LR}$  & $Z'_{\chi}$  & $Z'_{\eta}$  & $Z'_{\psi}$\tabularnewline
\hline
700  & 13.4  & 18.4  & 20.3  & 20.4  & 20.3\tabularnewline
\hline
1000  & 8.0  & 10.5  & 11.3  & 11.6  & 11.6\tabularnewline
\hline
1500  & 4.1  & 5.2  & 5.4  & 5.4  & 5.4\tabularnewline
\hline
2000  & 2.4  & 2.9  & 3.0  & 2.9  & 2.9\tabularnewline
\hline
3000  & 1.4  & 1.7  & 1.8  & 1.8  & 1.8\tabularnewline
\hline
\end{tabular}
\end{table}

\subsection{The top quark pair production}

\begin{figure}
\includegraphics[scale=0.6]{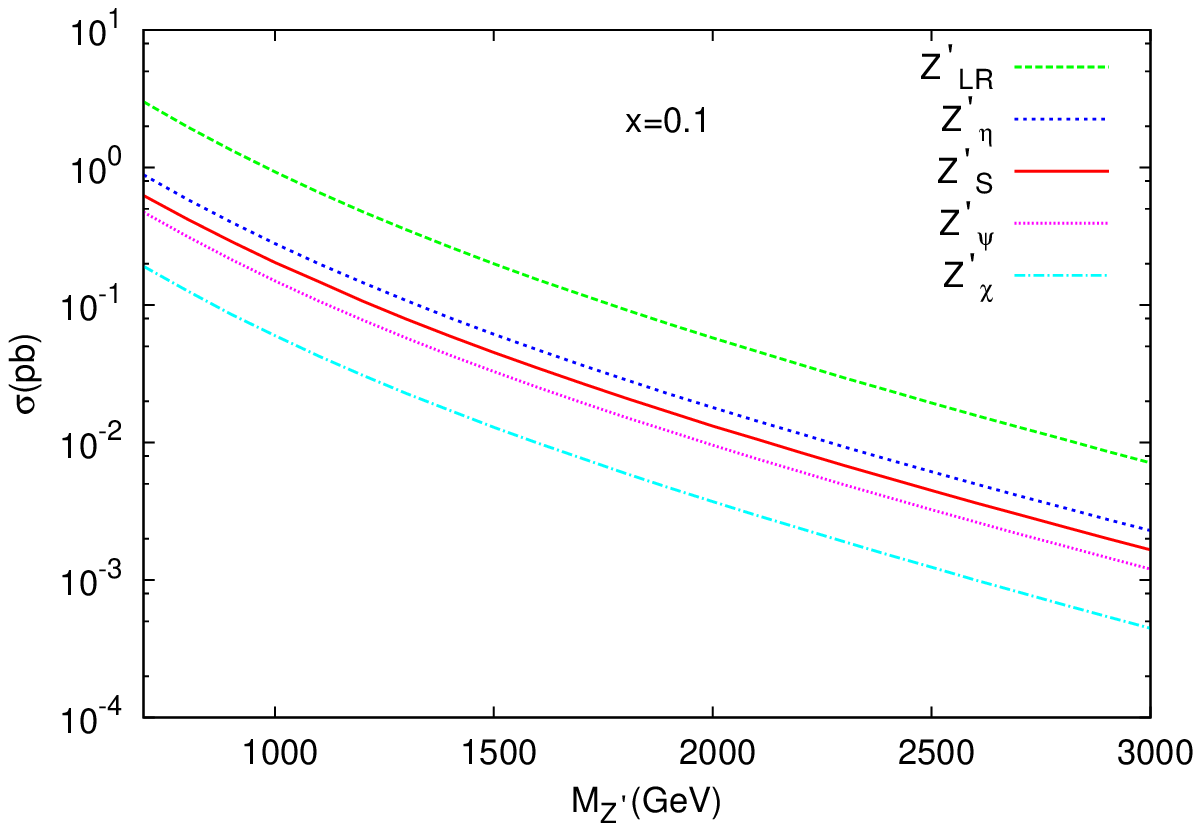} \includegraphics[scale=0.6]{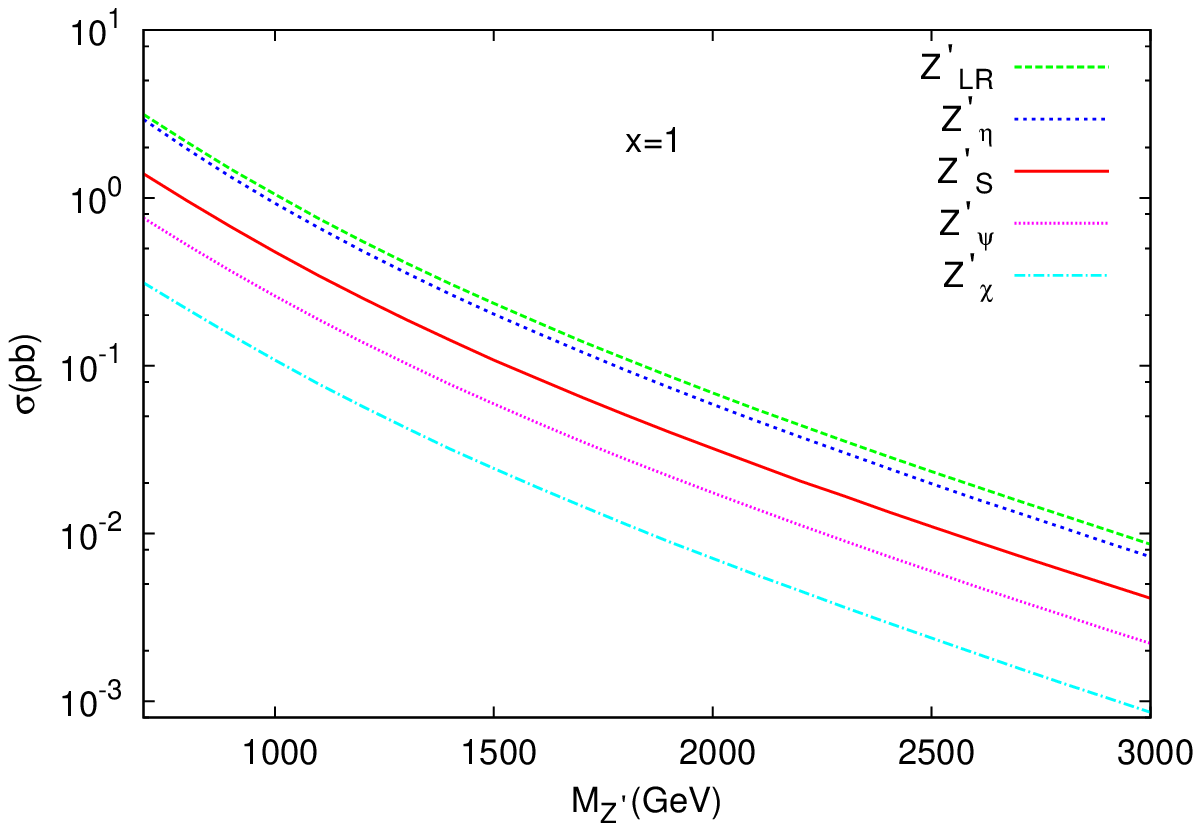}

\caption{The total cross sections for the top pair production at the LHC (with
$\sqrt{s}=14$ TeV) depending on the $Z'$ mass in the framework of
different models with $x=0.1$ and $x=1$. \label{fig:fig8}}

\end{figure}

\begin{figure}
\includegraphics[scale=0.8]{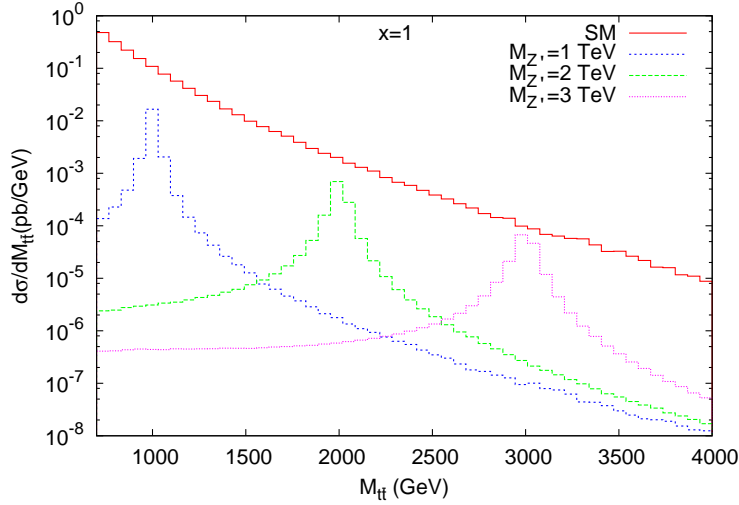}

\caption{The invariant mass distribution of the top pairs for the SM and sequential
$Z'$ model with $x=1$ at the LHC ($\sqrt{s}=14$ TeV). \label{fig:fig9} }

\end{figure}

\textit{\emph{We investigate the top pair production via $Z'$ boson
exchange at the LHC.}} The $Z'$ boson contributes in the $t$-channel
via the FCNC and in the $s$-channel through family diagonal neutral
current couplings with a strength scaled by the parameter $x$. \textit{\emph{The
total cross sections for the top pair production with $x=0.1$ and
$x=1$ for different models of $Z'$ at the LHC are plotted in Fig.}}\ref{fig:fig8}.
The main background has the same final state as the signal ($t\bar{t}$).
\textit{\emph{In the invariant mass analysis, we reconstruct the mass
of top pairs around the $Z'$ boson mass which are shown in Fig.}}
\ref{fig:fig9}. \textit{\emph{We assume $t(\bar{t})\to W^{+}(W^{-})b(\bar{b})$,
where the $W$ bosons decays}} leptonically. \textit{\emph{For each
of the $b$-quarks, we assume the $b$-tagging efficiency as $60\%$.}}
We calculate the cross section of the background in the mass bin widths
for each $M_{Z'}$ value; as an example, for the $M_{Z'}=1$ TeV we
take $\Delta M\simeq55$ GeV, and we find the background cross section
$\Delta\sigma_{B}=2.65\times10^{-1}$ pb for $pp\to W^{+}W^{-}b\bar{b}X$,
and $\Delta\sigma_{B}=3.97$ pb for $pp\to t\bar{t}X$ .

\textit{\emph{In Table \ref{tab:tab4}, we give the SS values for
$Z'$ boson in different models in the case of $x=0.1(1)$, considering
the}} leptonic channels of the top decays, taking $L_{int}=100$ fb$^{-1}$.

\begin{table}
\caption{The signal significance ($SS$) for the $Z'$ boson predicted in different
models, the $SS$ values show the case $x=0.1$ (1), with $L_{int}=100$ fb$^{-1}$. \label{tab:tab4}}

\begin{tabular}{|c|c|c|c|c|c|}
\hline
Mass (GeV)  & $Z'_{S}$  & $Z'_{LR}$  & $Z'_{\chi}$  & $Z'_{\eta}$  & $Z'_{\psi}$\tabularnewline
\hline
\hline
700  & 5.6(27.3)  & 27.0(25.5)  & 1.7(2.7)  & 7.9(12.1)  & 4.3(6.6)\tabularnewline
\hline
1000  & 3.6(17.6)  & 16.2(15.6)  & 1.0(1.8)  & 4.9(8.0)  & 2.6(4.4)\tabularnewline
\hline
2000  & 1.2(6.0)  & 5.3(5.2)  & 0.3(0.6)  & 1.6(2.8)  & 0.9(1.5)\tabularnewline
\hline
3000  & 0.5(2.6)  & 2.3(2.2)  & 0.1(0.3)  & 0.7(1.3)  & 0.4(0.7)\tabularnewline
\hline
\end{tabular}
\end{table}

\begin{figure}
\includegraphics[scale=0.8]{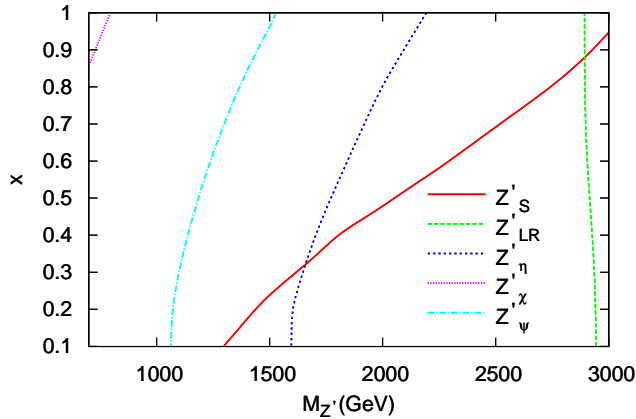}

\caption{The contour plot for the discovery of $Z'$ boson at the LHC ($\sqrt{s}=14$
TeV) with $L_{int}=150$ fb$^{-1}$. \label{fig:fig10}}

\end{figure}

We plot the $3\sigma$ contours in $(x-M_{Z'})$ plane, for different
$Z'$ models as shown in Fig. \ref{fig:fig10} at the LHC with $\sqrt{s}=14$
TeV and $L_{int}=150$ fb$^{-1}$. In order to cover all the interested $Z'$ models 
in the same range of parameter space, we take the integrated luminosity of 
$L_{int}=150$ fb$^{-1}$ in the leptonic $W$-decay channel.

\section{electron-positron collisions}

We anticipate that the LHC will explore the new physics directions
in the ongoing experiments, then the parameters of the new physics
will have been known with only moderate precision. The linear collider
at a high energy is expected to identify the model parameters and
make measurements with great precision beyond the hadron collider
reach.

In a lepton collider, the initial state is well known for both unpolarized
and polarized beams. The energy-momentum conservation can be used
in the full event reconstruction and the energy scan allow precise
measurement of the resonance parameters. For the $e^{+}e^{-}$ collider,
we consider the Compact Linear Collider (CLIC) in two beam acceleration
technology allowing the preferable center of mass energy $1(3)$ TeV
with $L=2(8)\times10^{34}$ cm$^{-2}$s$^{-1}$\cite{clic}.

\subsection{The single top quark production}

\begin{figure}
\includegraphics[scale=0.8]{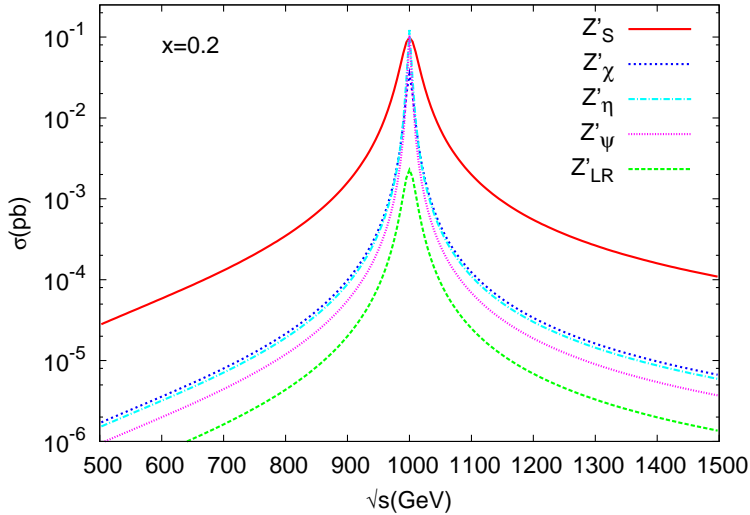}

\caption{The production cross sections for $e^{+}e^{-}\to t\bar{c}+\bar{t}c$
at the collision center of mass energy range 500-1500 GeV. Here, we
set the flavor changing parameter $x=0.2$. The curves show five models
of the $Z'$ boson with mass $M_{Z'}=1000$ GeV: sequential type (S),
left-right symmetrical model (LR), $E_{6}$-inspired models ($\chi,\psi,\eta$)
mentioned in the text. \label{fig:fig11}}

\end{figure}

\begin{figure}
\includegraphics[scale=0.8]{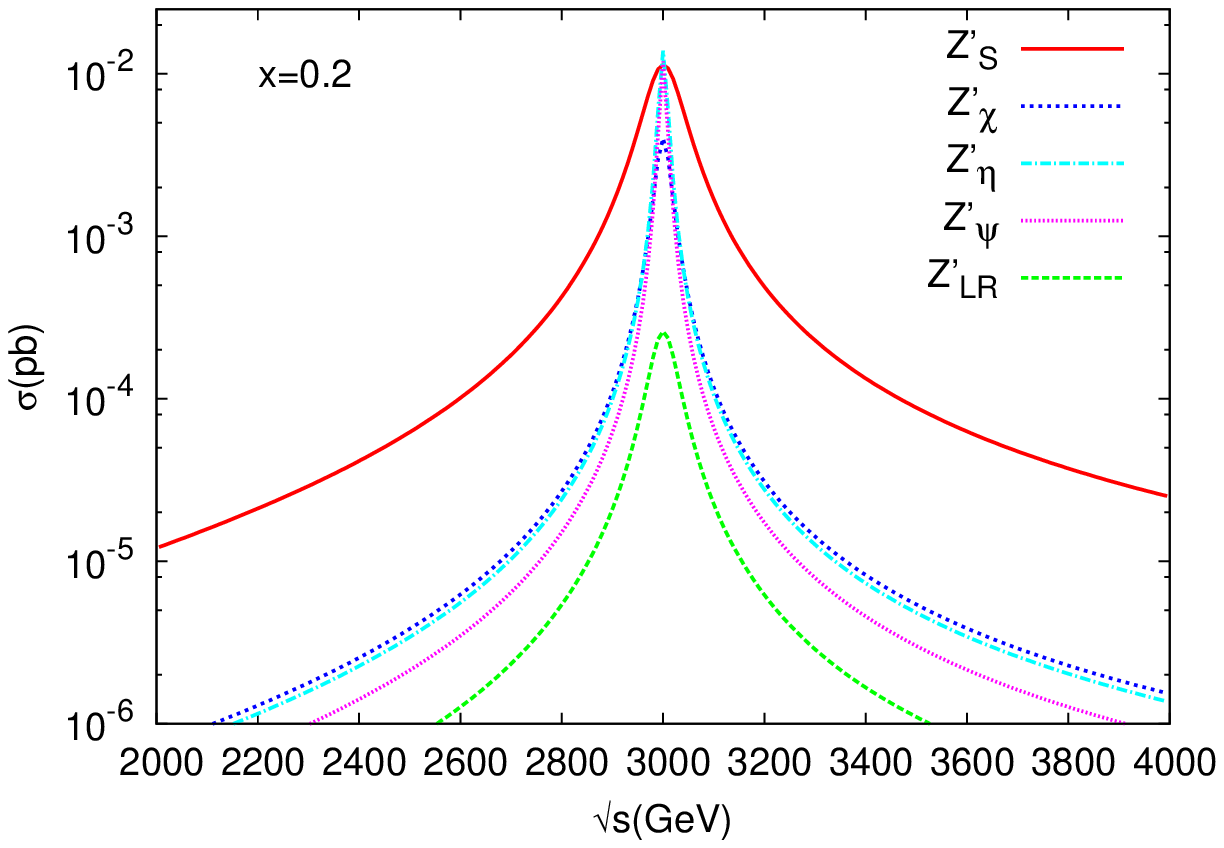}

\caption{The production cross sections for $e^{+}e^{-}\to t\bar{c}+\bar{t}c$
at the collision center of mass energy range 2000-4000 GeV. Here,
we set the $Z'$ bozon mass $M_{Z'}=3000$ GeV and the flavor changing
parameter $x=0.2$. The curves show five models of the $Z'$ boson:
sequential type (S), left-right symmetrical model (LR), $E_{6}$-inspired
models ($\chi,\psi,\eta$) mentioned in the text. \label{fig:fig12} }

\end{figure}

The production cross sections for $e^{+}e^{-}\to t\bar{c}+\bar{t}c$
are shown in Fig. \ref{fig:fig11} at the collision center of mass
energy ranging $0.5-1.5$ TeV with the assumption $M_{Z'}=1$ TeV. We also 
check our results for $x=0.1$ with the similar results given in \cite{arhrib06}.
In Fig. \ref{fig:fig12}, the cross sections around resonance $(M_{Z'}=3$
TeV) are shown for the center of mass energy range of 2-4 TeV. The
sequential model gives the largest cross sections, while the $Z'_{LR}$
model gives the lowest cross section around the resonance in the $e^{+}e^{-}$
collisions.

\begin{figure}
\includegraphics[scale=0.8]{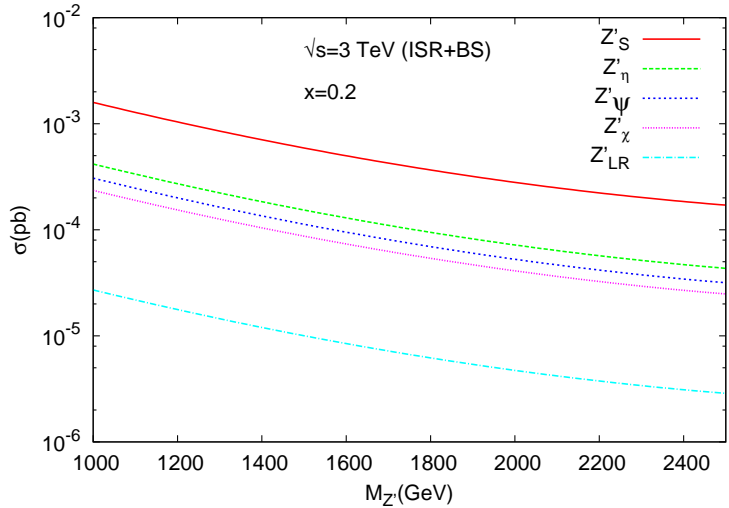}

\caption{The cross section for the process $e^{+}e^{-}\to
t\bar{c}+\bar{t}c$ depending on the $Z'$ mass at $\sqrt{s}=3$ TeV
with the ISR and BS effects. \label{fig:fig13}}

\end{figure}

A specific feature of the linear colliders is the presence of initial
state radiation (ISR) and beamstrahlung (BS). The effects of the ISR and BS 
can lead to a decrease
in the cross section. Being at the resonance for $M_{Z'}=3$ TeV,
we have the cross sections $9.04\times10^{-1}$ fb ($1.93\times10^{-2}$
fb) for sequential (LR) $Z'$ model with $x=0.2$. Taking center of
mass energy $\sqrt{s}=3$ TeV, and the mass range 1 TeV<$M_{Z'}$<2.5
TeV the signal cross sections with the ISR+BS effects are shown in
Fig. \ref{fig:fig13}. When calculating the ISR and BS effects, we
take updated beam parameters for the CLIC at $\sqrt{s}=$3 TeV as
follows: the beam sizes $\sigma_{x}+\sigma_{y}$= 46 nm, bunch length
$\sigma_{z}$= 44 $\mu$m, number of the particles in the bunch $N=3.72\times10^{9}$
\cite{clicnew}.

\begin{figure}
\includegraphics[scale=0.8]{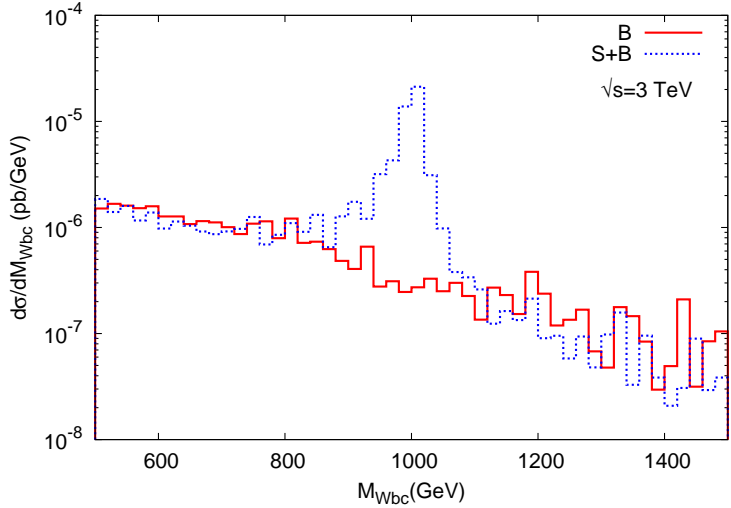}

\caption{The invariant mass distribution of the $W^{+}b\bar{c}$ system for
the sequantial model with $M_{Z'}=1$TeV. \label{fig:fig14}}

\end{figure}

We plot the invariant mass distributions for the $W^{+}b\bar{c}$
system in the final state, the signal has the peak around $M_{Z'}=1$
TeV over the smooth background as shown Fig. \ref{fig:fig14}. For
each model we can calculate the signal significance using signal and
background events in the chosen invariant mass intervals and different
mixing parameters $x$ as presented in Table \ref{tab:5}.

\begin{figure}
\includegraphics[scale=0.8]{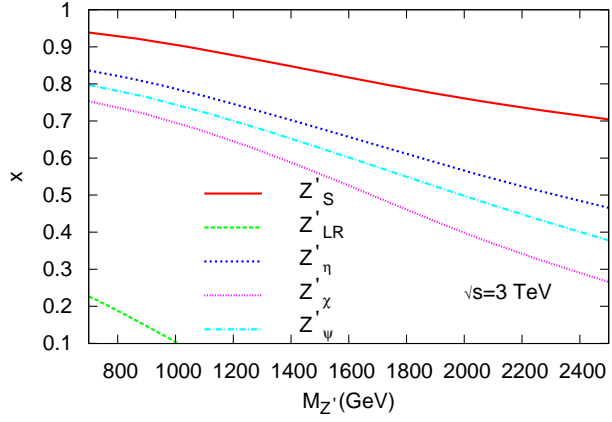}

\caption{The discovery region for the mass of the $Z'$ boson and the FCNC
mixing parameter for the single production of the top quarks. \label{fig:fig15}}

\end{figure}

We require the events lying in the invariant mass intervals $150$
GeV$<M_{Wb}<200$ GeV and
$M_{Z'}-2\Gamma_{Z'}<M_{Wbc}<M{}_{Z'}+2\Gamma_{Z'}$. Assuming
Poisson statistics we require $SS>3$ for signal observation, for
this case we can cover almost five $Z'$ models as shown in Fig.
\ref{fig:fig15}. Nevertheless, we need to gather more luminosity (at
least a factor of 10) to see more realization of the $LR$ model.

\begin{table}
\caption{The statistical significance for the $Z'$ search in the single top
production at the CLIC ($\sqrt{s}=3$ TeV) with $L_{int}=10^{5}$
pb$^{-1}$. The values are given for $x=0.1(0.6)$.\label{tab:5}}

\begin{tabular}{|c|c|c|c|c|c|}
\hline
Mass (GeV)  & $Z'_{S}$  & $Z'$$_{LR}$  & $Z'_{\chi}$  & $Z'_{\eta}$  & $Z'_{\psi}$\tabularnewline
\hline
\hline
700  & 11.31(4.5)  & 1.1(0.4)  & 4.1(1.6)  & 5.8(2.4)  & 4.9(2.0)\tabularnewline
\hline
1000  & 9.2(2.3)  & 0.9(0.3)  & 3.3(1.3)  & 4.7(1.9)  & 4.0(1.6)\tabularnewline
\hline
2000  & 3.9(1.6)  & 0.4(0.2)  & 1.4(0.6)  & 2.0(0.8)  & 1.7(0.7)\tabularnewline
\hline
3000  & 6.9(2.8)  & 0.8(0.3)  & 3.6(1.5)  & 6.7(2.7)  & 6.1(2.5)\tabularnewline
\hline
\end{tabular}
\end{table}

\subsection{The top quark pair production}

\begin{figure}
\includegraphics[scale=0.8]{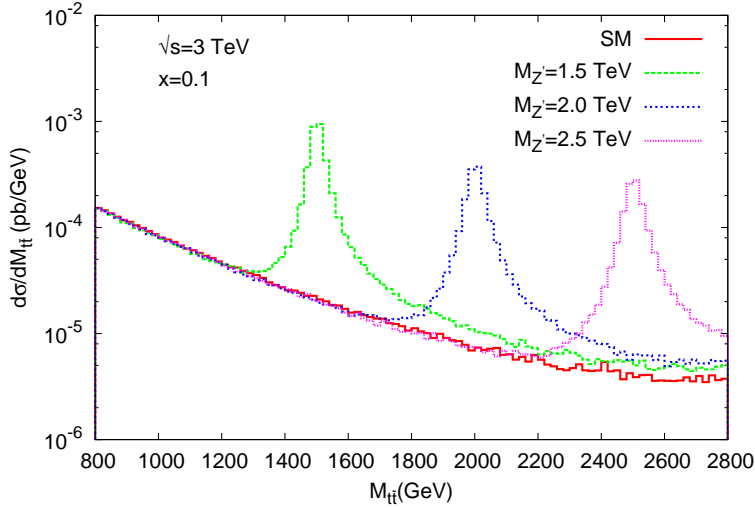}

\caption{The invariant mass distribution for top pairs in the sequential $Z'$
model at the CLIC with $\sqrt{s}=3$ TeV. \label{fig:fig16}}

\end{figure}

\begin{figure}
\includegraphics[scale=0.8]{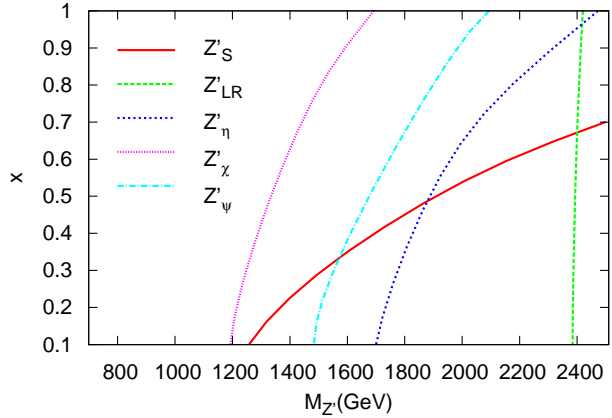}

\caption{The attainable region for the different $Z'$ models in the $x-m_{Z'}$
plane at the CLIC with $\sqrt{s}=3$ TeV and $L_{int}=400$ pb$^{-1}$.
\label{fig:fig17}}

\end{figure}

The $Z'$ boson will enhance the cross section of the pair production
of top quarks at the CLIC energies. Having the center of mass energy
$\sqrt{s}=3$ TeV, we plot the invariant mass distribution of the
background and signal (for $M_{Z'}=1.5$, $2$ and $2.5$ TeV) as shown
in Fig. \ref{fig:fig16}. Here, we consider the process
$e^{+}e^{-}\to t\bar{t}$ for both the background and signal. In the
analysis, we take into account leptonic decays of the top quarks
with the corresponding branchings and efficiency factors. We also
consider the background process $e^{+}e^{-}\to W^{+}W^{-}b\bar{b}$,
here one can apply the $Wb$ invariant mass cut (around top mass) to
reduce this background. We calculate the signal and background
events in the invariant mass intervals $\Delta m$ for an estimation
of the signal significance. Taking integrated luminosity
$L_{int}=10^{5}$ pb$^{-1}$ we calculate statistical significance for
the different $Z'$ mass points as shown in Table \ref{tab:6}. The
discovery region for $Z'$ searches is given in Fig. \ref{fig:fig17}.

\begin{table}
\caption{The statistical significance for the $Z'$ search in top pair production
at CLIC ($\sqrt{s}=3$TeV). The values are given for $x=0.1(1)$ \label{tab:6}.}

\begin{tabular}{|c|c|c|c|c|c|}
\hline
Mass (GeV)  & $Z'_{S}$  & $Z'$$_{LR}$  & $Z'_{\chi}$  & $Z'_{\eta}$  & $Z'_{\psi}$\tabularnewline
\hline
\hline
700  & 71.2(217.6)  & 161.2(157.3)  & 69.2(96.6)  & 112.0(153.3)  & 92.4(126.7)\tabularnewline
\hline
1000  & 53.8(181.7)  & 126.7(126.8)  & 155.3(82.3)  & 87.2(126.8)  & 73.7(107.7)\tabularnewline
\hline
2000  & 26.0(81.9)  & 55.2(56.0)  & 24.4(37.7)  & 37.6(56.7)  & 32.4(49.1)\tabularnewline
\hline
3000  & 50.1(151.4)  & 115.6(117.3)  & 67.0(101.0)  & 137.3(204.6)  & 122.2(183.4)\tabularnewline
\hline
\end{tabular}
\end{table}

\section{Analysis}

Heavy quark flavor tagging is useful to analyze the final state
event topology. One can identify a heavy flavor jet and measure the
invariant mass of the hadrons at a secondary vertex to differentiate
the charm and bottom jets. The charm quark hadronizes immediately
after it is produced. A charmed jet has a secondary vertex mass
ranging from 0 to 2 GeV with a peak around 1 GeV, while bottom jet
has the largest secondary vertex mass with a tail up to 4 GeV. The
light quark jets have the smallest secondary vertex masses. In
addition, the Monte Carlo samples can be used to determine the
fractions of charm, bottom and other light quarks in the event.
Here, we assume that one can have success for tagging the charmed
meson together with a single top quark. The FCNC single production
of top quarks through $Z'$ exchange can be probed at CLIC.
For the pair production of top quarks through the $Z'$ contribution,
we assumed the $b$-tagging efficiency as $60\%$ and we use leptonic
decay mode of the $W$-boson.

\section{Conclus\i{}on}

We find the discovery regions of the parameter space for the 
single and pair FCNC productions of top quarks via $Z'$ exchanges. 
The tree-level FCNC couplings of the top quark can emerge in the models
with an extra $U(1)$ group. In the models considered in this paper,
the single and pair production of top quarks at the LHC can have the
contributions from the couplings of $Z'q\bar{q}$ and the FCNC couplings
of $Z'q\bar{q}'$ (where $q,q'=u,c,t$). For a mixing parameter 
$x=0.2$ the LHC can discover top FCNC up to the mass $m_{Z'}=1.7-3$ TeV 
depending on the $Z'$ models. While at the CLIC, only the
$Z't\bar{t}$ interaction contributes to the process $e^{+}e^{-}\to t\bar{t}$
depending on the strength of parameter $x$. We can also investigate
the $Z't\bar{c}$ coupling at the linear colliders independent of
the other $Z'q\bar{q}'$ couplings. In case of the resonance production 
CLIC has better potential than the LHC in searching the  
single top FCNC via the $Z'$ boson.

\begin{acknowledgments}
O.C's work is partially supported by Turkish Atomic Energy Authority (TAEK)
under the grant No. CERN-A5.H2.P1.01-10. O.C's work is also partially 
supported by State Planning Organization (DPT) under
the grant No. DPT2006K-120470. I.T.C. acknowledges the support from
CERN Physics Department. 
\end{acknowledgments}

\end{document}